\documentclass[prd, floatfix, superscriptaddress, twocolumn]{revtex4}
\usepackage[dvips]{graphicx}
\usepackage{txfonts}

\begin{document}

\title{On the stability of self-gravitating accreting flows}
\author{Patryk Mach}
\affiliation{M. Smoluchowski Institute of Physics, Jagiellonian University, Reymonta 4, 30-059 Krak\'{o}w, Poland}
\author{Edward Malec}
\affiliation{M. Smoluchowski Institute of Physics, Jagiellonian University, Reymonta 4, 30-059 Krak\'{o}w, Poland}
\affiliation{Physics Department, UCC, Cork, Ireland}
\date{\today}

\begin{abstract}
Analytic  methods show stability of the stationary accretion of test fluids but they are inconclusive in the case 
of self-gravitating stationary flows. We investigate numerically stability of those stationary flows onto compact 
objects that are transsonic and rich in gas. In all studied examples solutions appear stable. Numerical 
investigation suggests also that the analogy between sonic and event horizons holds for small perturbations 
of compact support but fails in the case of finite perturbations.
\end{abstract}

\maketitle

\section{Introduction}

Investigation of the spherical accretion onto compact objects starts with seminal works of Hoyle, Lyttelton, and 
Bondi \cite{hoyle_lyttleton, lyttleton_hoyle, bondi_hoyle} describing the infall of dust matter onto a surface of 
a star moving through the interstellar medium. The first hydrodynamical analysis of spherical accretion was 
presented in 1952 by Bondi \cite{bondi} who considered a spherically symmetric flow of a polytropic perfect fluid 
in a Keplerian gravitational potential. A general-relativistic model of accretion in the Schwarzschild space-time 
has been developed by Michel \cite{michel}. In all these works the test fluid approximation has been adopted.

The first general-relativistic model with self-gravitating steady fluids has been analyzed by Malec \cite{malec}.
 This work has been later continued by Karkowski, Kinasiewicz, Mach, Malec, and \'{S}wierczy\'{n}ski \cite{kkmms}, 
and resulted in finding a whole family of steady transsonic solutions of Einstein equations describing 
a self-gravitating cloud of gas accreting onto a central compact object (a black hole in this particular case). 
The most striking fact about these solutions is that, given fixed asymptotic parameters of the model
 (such as the size of the cloud, speed of sound at the outer boundary, and the total asymptotic mass of the system), 
there exist two different transsonic solutions corresponding to the same accretion rate---one, for which most 
of the mass is contained in the central object, and the other, where the amount of mass contained in 
the accreting fluid constitute almost all the mass of the entire configuration. The first class 
of the solutions contains a subset of the test fluid flows found by Michel. Other solutions are new. 
Let us stress, that although these solutions are not available in a closed form, 
many of their parameters and properties can be inferred by analytical means. Such flows might be associated with the 
Thorne--\.Zytkow stars \cite{thorne, thorne2} or quasistars \cite{begelman_et_al}.

We should also mention here the work of Papadopoulos and Font \cite{papadopoulos_font}, 
who constructed a general-relativistic hydrodynamical code capable of simulating self-gravitating flows 
and applied it to the strongly perturbed Michel's solution. The fluid perturbation was treated as 
self-gravitating but the background solution corresponded to the test fluid regime. A similar 
investigation, but in the context of radiation hydrodynamics, has been done by Zampieri et al. \cite{Zampieri}. 
They investigated the stability of solutions in Schwarzschild space-time found by Nobili et al. \cite{Nobili}.

The first proof of the stability of the transsonic accretion in the Newtonian and relativistic, spherically 
symmetric cases, given by Moncrief \cite{moncrief}, was restricted to the test fluid approximation. Analytic 
methods are inconclusive in the case of self-gravitating accretion flows \cite{ladek}. Below we report
 the results of a numerical analysis of the stability of both branches of solutions found in \cite{kkmms}. 
An interesting by-product of this investigation is that in the nonlinear regime sonic horizons are movable 
and the signal can get out from within the sonic horizon in the original steady flow. That hints to 
the limited validity of the formal analogy between sonic and event horizons. 

A stability analysis of Newtonian accretion solutions  with self-gravitating flows has been done for axially 
symmetric perturbations. In all examined cases we have observed the stable behavior of both aforementioned 
branches of solutions: those corresponding to the test fluid as well as those where the mass of the fluid 
dominates over the mass of a central object.

The order of the forthcoming sections is as follows. Section II gives a short description of the two branches 
of accreting flows in spherically symmetric space-times. Section III introduces the dynamical equations 
of motion in the form adopted in the numerical code. Section VI reports results concerning stability 
of accreting flows. The next section shows that sonic horizons can be penetrated from within by large perturbations. 
This suggests the limited validity of the analogy between sonic and event horizons. The stability 
of Newtonian accretion under axisymmetric perturbations is shown in section VIII. Obtained results 
are briefly reviewed in section VIII.

\section{Steady solutions}

A general spherically symmetric space-time can be described by the line element
\begin{equation}
\label{comoving_metric}
ds^2 = -N^2 d {\tilde t}^2 + \alpha dr^2 + R^2 \left(d\theta^2 + \sin^2 \theta d\phi^2 \right),
\end{equation}
where $N$, $\alpha$, and $R$ are functions of the coordinate radius $r$ and the asymptotic time variable $\tilde t$.

The extrinsic curvature of a slice of constant time $\tilde t$ of the space-time with the metric given by 
(\ref{comoving_metric}) has the following non-zero elements: $K_r^r = 
\partial_{\tilde t} \alpha / (2 \alpha N)$, $K_\theta^\theta = K_\phi^\phi = \partial_{\tilde t} R /(RN)$. 
Accordingly the trace of the extrinsic curvature can be written as $\mathrm{tr} K = N^{-1} \partial_{\tilde t} \ln 
\left( \sqrt{\alpha} R^2 \right)$.

Similar calculations can be performed for the two-spheres of constant radius $r$ embedded in a given temporal slice.
 The result for the trace of the extrinsic curvature (twice the mean curvature of the surface) is 
$k = 2 \partial_r R / \left( R \sqrt{\alpha} \right)$.

In what follows we will consider the evolution of a spherical cloud of perfect fluid accreting onto 
a central object and described by the energy-momentum tensor
\begin{equation}
\label{energy_momentum_tensor}
T^{\mu\nu} = \left(p + \rho \right) u^\mu u^\nu + p g^{\mu \nu},
\end{equation}
where $p$ is the pressure, $\rho$ the energy density, and $u^\mu$ the four-velocity of the fluid.

We start working with the comoving gauge, so that $u^r = u^\theta = u^\phi = 0$ 
(there exist a suitable geometric condition imposed on extrinsic curvatures $K_i^j$ for such a choice of coordinates 
\cite{malec}), and introduce a function $U \equiv RK_\theta ^\theta = \partial_{\tilde t} R/N $.
 It has the meaning of the spatial part of the fluid four-velocity computed in the reference frame
 $(t^\prime, r^\prime)$ that has been obtained by the transformation $(\tilde t, r) \mapsto (t^\prime = \tilde t, r^\prime 
= R(\tilde t,r))$. We also introduce the quasilocal mass, which can be easily expressed as
\[ m(R) = m_\mathrm{tot} - 4 \pi \int_R^\infty {R^\prime}^2 \rho d R^\prime. \]
Here $m_\mathrm{tot}$ is the total asymptotic mass of the configuration, and $R_\infty$ denotes the size 
of the accretion cloud. Another important quantity, the local speed of sound $a$ is defined by
\[ a^2 = \frac{1}{h} \left( \chi + \frac{p \kappa}{n^2} \right). \]
Symbols $\chi$ and $\kappa$ are used here to denote derivatives
\[ \chi = \left( \frac{\partial p}{\partial n}\right)_\epsilon, \;\;\; \kappa = 
\left( \frac{\partial p}{\partial \epsilon} \right)_n, \]
which have to be computed according to the assumed equation of state; the quantity 
$h = (\rho + p)/n$ is the specific enthalpy. For a barotropic equation of state the above definition reduces to 
$a^2 = dp / d \rho$.

\begin{figure}
\begin{center}
\includegraphics[width=80mm]{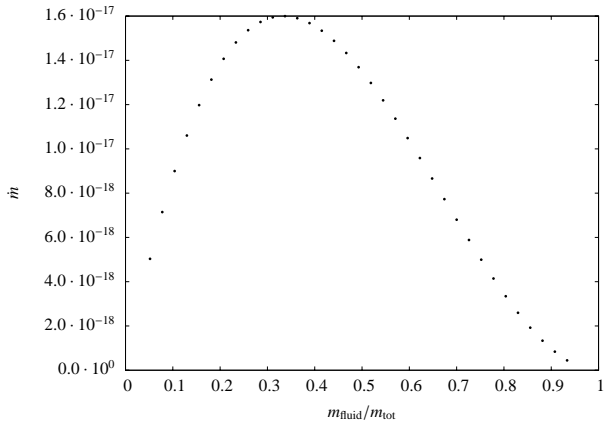}
\end{center}
\caption{The dependence of the accretion rate $\dot m$ on the $m_\mathrm{fluid}$. Here $m_\mathrm{tot} = 1$.}
\label{m_dot_graph}
\end{figure}

We will search for the solutions of the Einstein equations according to the following notion of 
stationarity. The accretion rate $\dot m = \partial_{t^\prime} m$, computed at a given areal 
radius $R$ should be constant in time. Similarly, other hydrodynamical quantities like 
the fluid velocity $U$, pressure $p$, energy density $\rho$, etc. should satisfy $\partial_{t^\prime} X = 0$, 
where $X = U, p, \rho, \dots$ These assumptions can be satisfied only approximately as the accreting 
fluid contributes to the growing mass of the central object and the whole configuration must change 
in time. We will, however, see that they lead to solutions characterized by very small accretion rates 
both in the test and in the heavy fluid regime. Thus, for a very long time (much larger than 
a characteristic dynamical time understood as a time required by a sound wave to travel across 
the cloud) the motion of the accreting gas remains almost unchanged, and the above assumptions 
are justified a posteriori.

Under these assumptions the set of Einstein partial differential equations and the equations 
expressing the conservation of the energy-momentum tensor reduces to a set of ordinary ones, namely
\[ \frac{d}{dR} \ln \left( \frac{N}{kR} \right) = \frac{16 \pi}{k^2 R} (\rho + p), \;\;\;  U = \frac{A}{R^2 n}, \]
\begin{equation}
N = \frac{Bn}{\rho + p}, \;\;\; kR = 2 \sqrt{1 - \frac{2m}{R} + U^2}.
\label{steady_eqns}
\end{equation}

Here $A$ and $B$ are for integration constants. The baryonic density $n$ is defined as a function 
assuring that $\nabla_\mu (n u^\mu) = 0$. When deriving these equations we have tacitly assumed 
that the equation of state is of a barotropic form $p = p(\rho)$ or, equivalently, $p = p(n)$. 
In the following we will specialize to the polytropic equation of state $p = Kn^\Gamma$.

In addition to these equations a suitable set of boundary conditions has to be specified. 
As such we usually choose the size of the accretion cloud $R_\infty$, its total mass $m_\mathrm{tot}$, 
the asymptotic value of the baryonic density $n_\infty$ (alternatively $\rho_\infty$ can be also used), 
and the asymptotic value of the speed of sound $a_\infty$.

Such boundary conditions still provide a family of the solutions parametrized by the value of 
$U_\infty$, or the constant $A$. In this work we are only interested in the so-called transsonic solutions---those 
for which far away from the central object the fluid is subsonic while in the central parts of the cloud it 
falls supersonically. Such a solution passes through the so-called sonic point, defined as a location where $2U/(kR) = a$.

The analysis presented in \cite{kkmms} is concerned with accretion flows onto a central black hole which 
will naturally appear in the model if we continue to integrate the equations from the outer boundary inwards 
until the apparent horizon is passed by. In the terms of this paper the apparent horizon is defined as 
a surface on which the optical scalar $\theta_+ = kR/2 + U$ vanishes. According to this definition, the radius 
of the apparent horizon $R_\mathrm{BH}$ and the mass $m_\mathrm{BH} = m(R_\mathrm{BH})$ (the mass of the black 
hole) satisfy the standard relation $R_\mathrm{BH} = 2 m_\mathrm{BH}$. Apart from the mass of the black hole 
we also define the fluid mass as $m_\mathrm{fluid} = m_\mathrm{tot} - m_\mathrm{BH}$.

Fig.\,\ref{m_dot_graph} shows the dependence of the accretion rate on the ratio of $m_\mathrm{fluid}/m_\mathrm{tot}$ 
for the specific case of a sequence of polytropic models with $R_\infty = 10^6$, $m_\mathrm{tot} = 1$ and $a_\infty = 0.1$. 
(The ratio of $m_\mathrm{fluid}/m_\mathrm{tot}$ scales linearily with $n_\infty$.)  Clearly, the slowly accreting 
regime corresponds either to a situation with a very small mass of the fluid as compared to the mass of 
the central black hole or to converse case where almost entire mass is in the form of fluid. 
This can actually be proved analytically \cite{kkmms}. Moreover, it can be shown that the maximum 
of $\dot m$ corresponds to $m_\mathrm{fluid}/m_\mathrm{tot} = 1/3$.

The attempts to investigate the stability of the new branch of massive solutions analytically have 
failed to give any conclusive results (see, e.g., \cite{ladek}). Due to this fact we devote this paper 
to study this numerically.

\section{Description of the numerical code}

The dynamical code used to investigate the stability of steady solutions was constructed in a similar fashion 
to the one described in \cite{romero_1996}. It is a modern version of a High Resolution Shock Capturing (HRSC) 
scheme based on the Godunov-type methods developed for solving the equations of hydrodynamics.

For the construction of the code the polar gauge has been used. The metric is assumed to be of the form
\begin{equation}
\label{polar_metric}
ds^2 = - \alpha^2 dt^2 + X^2 dR^2 + R^2 \left( d \theta^2 + \sin^2 \theta d \phi^2 \right),
\end{equation}
where the lapse $\alpha$ and $X$ are functions of the areal radius $R$ and time $t$. Following \cite{banyuls_et_al} 
we introduce the Lorentz factor $W = \alpha u^t$ and the three-velocity $v^i = u^i/W$. The equations of conservation 
of the energy-momentum tensor $\nabla_\mu T^{\mu\nu} = 0$ and the baryonic density $\nabla_\mu(n u^\mu) = 0$ can 
be now written as
\[ \partial_t \mathbf q + \frac{1}{XR^2} \partial_R \left( \alpha X R^2 \mathbf F \right) = \alpha \mathbf \Sigma - 
\left( \partial_t \ln X \right) \mathbf q. \]
Here $\mathbf q$ denotes a vector of conserved quantities
\[ \mathbf q = \left( D, S, \tau \right)^T = \left( n W, n h W^2 v_R, n W (h W - 1) - p \right)^T, \]
$\mathbf F$ stands for the flux vector
\[ \mathbf F = \left( n W v^R, n h W^2 v_R v^R + p, n W (h W - 1) v^R \right)^T, \]
and $\mathbf \Sigma$ denotes the source terms
\[ \mathbf \Sigma = \left( \begin{array}{c}
0 \\
\left( n h W^2 v_R v^R + p \right) \frac{\partial_R X}{X} - \left( n h W^2 - p \right) \frac{\partial_R \alpha}{\alpha} 
+ \frac{2 p}{R} \\
- n h W^2 v^R \frac{\partial_R \alpha}{\alpha} - \left( n h W^2 v_R v^R + p \right) \frac{\partial_t X}{\alpha X}
\end{array} \right). \]

Time derivatives of conserved quantities $\mathbf q$ are computed according to the following version 
of the method of lines
\begin{eqnarray}
\left( \frac{d\mathbf q}{dt} \right)_i & = & - \frac{(\alpha X R^2 \hat \mathbf F)_{i+1/2} - (\alpha X R^2 
\hat \mathbf F)_{i-1/2} }{X_i R_i^2 \Delta R_i} + \nonumber \\
&& +  \left( \alpha \mathbf \Sigma - \frac{\partial_t X}{X} \mathbf q \right)_i,
\label{update_scheme}
\end{eqnarray}
where lower indices refer to spatial cells (shells of constant radius). Values of $\mathbf q$ 
corresponding to the subsequent time-step are obtained using standard Runge--Kutta methods.

The numerical scheme is stabilized by a suitable choice of the numerical fluxes $\hat \mathbf F_{i+1/2}$. 
In most of the modern HRSC schemes numerical fluxes at the cells interfaces are computed based on 
the solutions to the local Riemann problems that arise naturally between each of the cells' interfaces. 
In the following $\mathbf q_{\mathrm{L},i+1/2}$ and $\mathbf q_{\mathrm{R},i+1/2}$ will denote left and right 
Riemann states at the $i + 1/2$ interface. In order to provide higher order of the spatial accuracy the states 
$\mathbf q_{\mathrm{L},i+1/2}$ and $\mathbf q_{\mathrm{R},i+1/2}$ are computed as follows
\begin{eqnarray*}
\mathbf q_{i + 1/2}^\mathrm{L} & = & \mathbf q_i + \mathbf S_i \left( R_{i + 1/2} - R_i \right), \\
\mathbf q_{i + 1/2}^\mathrm{R} & = & \mathbf q_{i + 1} + \mathbf S_{i + 1} \left( R_{i + 1/2} - R_{i + 1} \right).
\end{eqnarray*}
Here $R_i$ and $R_{i+1/2}$ are the positions of the cells' centers and interfaces respectively. 
The slope limiters $S_i$ are defined by
\[ \mathbf S_i = \mathrm{minmod} \left( \frac{\mathbf q_{i+1} - \mathbf q_i}{R_{i+1} - R_i}, \frac{\mathbf q_i - 
\mathbf q_{i - 1}}{R_i - R_{i-1}} \right), \]
where the ``minmod'' function has been introduced as in \cite{van_leer}
\[ \mathrm{minmod}(a,b) = \left\{ \begin{array}{ll} a & \mathrm{if} \; |a| < |b|, \; ab > 0, \\b & \mathrm{if} \;
 |a| > |b|, \; ab > 0, \\0 & \mathrm{if} \; ab \leqslant 0. \end{array} \right. \]

As a method to compute numerical fluxes $\hat \mathbf F_{i+1/2}$ we have used two versions of a scheme 
proposed originally by Donat and Maraquina in \cite{donat_maraquina}, both of them being based on 
the spectral decomposition of the Jacobian $\partial \mathbf F / \partial \mathbf q$ consisting 
of three eigenvalues $\lambda_p$, left eigenvectors $\mathbf l_p$ and right ones $\mathbf r_p$. 
In the original version of the algorithm one starts by computing the following variables
\[ \omega^p_\mathrm{L} = \mathbf l^p_\mathrm{L} \cdot \mathbf q_\mathrm{L}, \;\;\; \omega^p_\mathrm{R} = 
\mathbf l^p_\mathrm{R} \cdot \mathbf q_\mathrm{R} \]
and
\[ \phi^p_\mathrm{L} = \mathbf l^p_\mathrm{L} \cdot \mathbf F (\mathbf q_\mathrm{L}), \;\;\; \phi^p_\mathrm{R}
 = \mathbf l^p_\mathrm{R} \cdot \mathbf F (\mathbf q_\mathrm{R}), \]
where $p$ numbers the eigenvectors of $\partial \mathbf F / \partial \mathbf q$. Now, for each $p$, 
the signs of $\lambda^p_\mathrm{L}$ and $\lambda^p_\mathrm{R}$ are inspected. If both eigenvalues 
$\lambda^p_\mathrm{L}$ and $\lambda^p_\mathrm{R}$ are positive, we define
\[ \phi_+^p = \phi^p_\mathrm{L}, \;\;\; \phi_-^p = 0, \]
while for both $\lambda^p_\mathrm{L}$ and $\lambda^p_\mathrm{R}$ having negative values, we set
\[ \phi_+^p = 0, \;\;\; \phi_-^p = \phi^p_\mathrm{R}. \]
If the signs of the two eigenvalues $\lambda^p_\mathrm{L}$ and $\lambda^p_\mathrm{R}$ differ, one 
defines $|\lambda^p|_\mathrm{max(L,P)} = \mathrm{max} \left( |\lambda^p_\mathrm{L}|, |\lambda^p_\mathrm{R}| \right)$ and
\begin{eqnarray*}
\phi_+^p & = & \frac{1}{2} \left( \phi^p_\mathrm{L} + |\lambda^p|_\mathrm{max(L,P)} \omega^p_\mathrm{L} \right), \\
\phi_-^p & = & \frac{1}{2} \left( \phi^p_\mathrm{R} + |\lambda^p|_\mathrm{max(L,P)} \omega^p_\mathrm{R} \right).
\end{eqnarray*}
The total numerical flux is now computed according to the formula
\[ \hat \mathbf F_{i + 1/2} = \sum_p \left( \phi^p_+ \mathbf r^p_\mathrm{L} + \phi_-^p \mathbf r^p_\mathrm{R} \right). \]

Apart from this algorithm, known in the literature as the Maraquina's flux formula, we have also implemented 
its slight modification described in \cite{genesis}. Here the numerical fluxes are computed as
\begin{eqnarray*}
\hat \mathbf F_{i+1/2} & = &  \frac{1}{2} \bigg\{ \mathbf F \left(\mathbf q_\mathrm{L} \right) + 
\mathbf F \left(\mathbf q_\mathrm{R} \right) + \\
&& - \sum_p \left| \lambda_p \right|_{\mathrm{max}(L,R)} \left( \left(\mathbf l_{p,\mathrm{R}} \cdot 
 \mathbf q_\mathrm{R} \right) \mathbf r_{p,\mathrm{R}}  -  \left(\mathbf l_{p,\mathrm{L}} \cdot 
\mathbf q_\mathrm{L} \right) \mathbf r_{p,\mathrm{L}} \right)  \bigg\}.
\end{eqnarray*}
We have not observed any significant difference in the code performance between these two schemes.

A spectral decomposition of the Jacobian $\partial \mathbf F / \partial \mathbf q$ can be easily 
obtained analytically. Its eigenvalues read
\[ \lambda_0 = v^R, \;\;\; \lambda_+ = \frac{X v^R + a}{X + a v_R}, \;\;\; \lambda_- = \frac{X v^R - a}{X - a v_R}. \]
Let us introduce the following quantities
\[ \mathcal{K} = \frac{\kappa}{\kappa - n a^2}, \;\;\; \mathcal{A}_\pm = \frac{1 - v_R v^R}{1 - v_R \lambda_\pm}, \]
and
\[ \Delta = \left( \mathcal{A}_+ \lambda_+ - \mathcal{A}_- \lambda_- \right) X^2 h^3 W 
\left( \mathcal{K} - 1 \right) \left( 1 - v_R v^R \right). \]
With their help the right eigenvectors of $\partial \mathbf F / \partial \mathbf q$ can be written as
\[ \mathbf r_0 = \left( \frac{\mathcal K}{hW}, v_R, 1 - \frac{\mathcal K}{hW} \right)^T, \]
\[ \mathbf r_\pm = \left(1, h W X^2 \lambda_\pm \mathcal A_\pm, h W \mathcal A_\pm - 1 \right)^T.\]
For the left eigenvectors we have
\[ \mathbf l_0 = \frac{W}{\mathcal K - 1} \left( h - W, W v^R, -W \right)^T, \]
\[ \mathbf l_\pm = \mp \frac{h^2}{\Delta} \left( \begin{array}{c}
h W \mathcal A_\mp \left( v_R - X^2 \lambda_\mp \right) - v_R + \mathcal K X^2 \mathcal A_\mp \lambda_\mp \\
1 - \mathcal K \mathcal A_\mp, \\
- v_R + \mathcal K \mathcal A_\mp X^2 \lambda_\mp
\end{array} \right). \]

In the spherically symmetric case and the polar gauge the Einstein equations can be reduced to just 
two ordinary differential equations in $R$, that can be subsequently solved by quadratures provided 
that the hydrodynamical equations are known (see e.g. \cite{gourgoulhon}). The first of these equations
\[ \partial_R m = 4 \pi R^2 \left( n h W^2 - p \right) = 4 \pi R^2 (D + \tau) \]
gives the radial derivative of the quasilocal mass, related to the metric function $X$ by
\begin{equation}
X = \frac{1}{\sqrt{ 1 - \frac{2m}{R}}}.
\label{X_function}
\end{equation}
The second one provides the radial derivative of the logarithm of the lapse
\begin{eqnarray*}
\partial_R \ln \alpha & = & X^2 \left( \frac{m}{R^2} + 4 \pi R \left( n h W^2 v_R v^R + p \right) \right) \\
& = & X^2 \left( \frac{m}{R^2} + 4 \pi R \left( S v^R + p \right) \right).
\end{eqnarray*}
These equations are integrated (numerically) at each time-step, that is after new values of 
the hydrodynamic quantities have been obtained. Notice that the equation for mass $m$ has to 
be solved before we attempt to integrate the equation for $\ln \alpha$. For a closed system 
one can fix the total mass of the system and integrate the first equation starting from 
the outer boundary $R_\infty$. Then the second equation can be integrated in the same way assuming, for 
instance, that the lapse at the outer boundary is given by the standard expression known from 
the Schwarzschild solution in the polar gauge.
\[ \alpha(R_\infty) = \sqrt{1 - \frac{2 m(R_\infty)}{R_\infty}}. \]
Let us remark, however, that such assumption for the lapse is not necessary for the proper behavior 
of the hydrodynamic part of the code.

The Einstein equations yield also the following expression
\[ \partial_t X = - 4 \pi n h \alpha W^2 X v_R R, \]
which is required in order to establish the source terms needed by the evolution scheme (\ref{update_scheme}).

Recovery of the primitive quantities $(n, v^R, p)$ from the conserved ones $(D, S, \tau)$ is performed 
every time-step by means of the Newton--Raphson technique (we solve an equation for the pressure $p$).

\section{Code tests}

Our numerical code has been tested on the spherical shock reflection problem \cite{blandford}, the Michel 
solution for spherical accretion in the Schwarzschild space-time \cite{michel}, and models of spherical 
polytropic stars \cite{tooper}.

The first test checks the validity of the hydrodynamical part of the code in spherical symmetry. Initial 
data for this problem consist of a spherically symmetric flow of perfect fluid with a constant, negative 
radial velocity $v_0$. The initial baryonic density distribution is also constant and the internal energy 
density is set to a negligibly small value. Such initial data evolve by producing a strong shock wave 
appearing at $r=0$ and propagating outward with velocity equal to
\[ v_s = \frac{(\Gamma - 1)W_0|v_0|}{W_0+1}, \]
where $W_0$ is the Lorentz factor corresponding to $v_0$.

The second test checks the validity of the implementation in a case of the test fluid accretion occurring 
in the fixed Schwarzschild background. The initial data for this test can be easily obtained by solving 
an algebraic equation for each value of the areal radius.

The initial data for the third test, namely a static solution describing a polytropic star, have to be 
computed by solving the Tolman--Oppenheimer--Volkov equations for a polytropic equation of state 
\cite{tooper}. In this case satisfactory results have been obtained using a Runge--Kutta scheme of 8-th 
order by Hairer and Wanner \cite{hairer}.

All these tests have been passed as desired, convincing us that all parts of the code work properly.

\section{Initial conditions}

In order to construct the initial data for the main study of this paper we have taken numerical, 
transsonic solutions to the equations (\ref{steady_eqns}) and added a perturbation in velocity. 
The way of obtaining such solutions is simple but not entirely obvious; we will describe it briefly here.

Instead of solving the equations in $R$ we introduce a new independent variable $\zeta = 1/R$, 
which results in a grid that becomes naturally dense in the inner regions of the cloud 
and relatively coarse outside. Next we introduce the following set of dependent variables
\begin{eqnarray*}
y_1 & = & \int_R^{R_\infty} dR^\prime {R^\prime}^2 \rho, \\
y_2 & = & 16 \pi \int_R^{R_\infty} \frac{\rho + p}{k^2R^\prime} dR^\prime, \\
y_3 & = & a^2,
\end{eqnarray*}
and express the equations (\ref{steady_eqns}) in terms of $y_1$, $y_2$, $y_3$, and $\zeta$. This yields 
a set of three differential--algebraic equations that can be integrated starting from $\zeta_\infty = 1/R_\infty$ 
towards increasing values of $\zeta$ (that is from the outer boundary to the center of the cloud) provided 
that the values of $m_\mathrm{tot}$, $n_\infty$, $a_\infty$ and $A$ are specified.

We solved these equations using \textsc{Daspk}---a solver for a system of differential and algebraic equations 
developed by Petzold, Brown, Hindermarsh, and Li \cite{brown_et_al, li_petzold}.

Usually, the solution found in this way will not pass through the sonic point. We can, however, search for 
the transsonic solution by exploiting the fact that it can be integrated to values of $\zeta$ corresponding 
to the region inside the apparent horizon. (This is not true for other solutions which break down outside 
the horizon.) Thus, in order to find a transsonic flow, we have implemented a bisection method, which 
looks for a value of $A$ giving the maximal $\zeta$ at which the corresponding solution crashes. 
After the proper value of $A$ had been obtained, we could confirm that the appropriate solution indeed passes 
through the sonic point.

Such a solution is expressed in the coordinate system where the time coordinate is the comoving time. Before 
treating it as a possible initial data one has to express the fluid velocity in the polar gauge. By adopting 
the areal radius as the radial coordinate we have changed from the comoving gauge metric given by the line 
element (\ref{comoving_metric}) to the coordinates $t^\prime (\tilde t,r) = \tilde t$, $r^\prime (\tilde t,r)
 = R(\tilde t,r)$. This leads to the line element
\begin{eqnarray*}
ds^2 & = & - \left( N^2 - \left( \frac{2}{kR}UN \right)^2 \right) {d t^\prime}^2 - 2UN \left( \frac{2}{kR} 
\right)^2 d t^\prime dR + \\
&& + \left( \frac{2}{kR} \right)^2 dR^2 + R^2 \left(d\theta^2 + \sin^2 \theta d\phi^2 \right).
\end{eqnarray*}
Here the four-velocity of the fluid reads
\[ u^{t^\prime} = \frac{1}{N}, \;\;\; u^R = U, \;\;\; u^\theta = u^\phi = 0. \]
The transition to the polar coordinates with the metric of the form (\ref{polar_metric}) can also be easily done. 
The function $X$ is given by (\ref{X_function}) and the velocity $v^R = u^R/W$---which is one of the dynamical 
variables in our code---can be written as
\[ v^R = \frac{U}{\sqrt{1 + X^2 U^2}}. \]
In this way we have obtained four functions $m(R)$, $v^R(R)$, $n(R)$, and $\epsilon(R)$ that correspond to 
stationary flow.  

\section{Stability of accreting self-gravitating flows}

The code described in the preceding section has been used in order to evolve perturbed steady accretion 
flows. 

In addition to the initial data one has to specify suitable boundary conditions. At the inner boundary 
outflow conditions have been assumed (the fluid was allowed to fall inward). This boundary was positioned 
outside the apparent horizon but always in the supersonic zone of the accretion cloud, so that the outflow 
conditions could be easily implemented. Thus, our numerical models resemble also a situation in which 
the accreting fluid is glued to the solid surface of the central body. Such an approach has been adopted 
for instance in \cite{karkowski_malec_roszkowski}, where the processes of the radiation transport through 
the accreting medium are also taken into account.

The outer boundary was kept fixed using the values obtained from the initial solution (the ghost zones 
were filled with the appropriate initial values). This last condition cannot be easily relaxed. After 
setting the outer boundary condition to an outflowing one, even the test fluid solution can be unstable.

As the first step in the stability analysis we performed a consistency check, by assuming initial data for 
the dynamical equations to be equal to $m(R)$, $v^R(R)$, $n(R)$, and $\epsilon(R)$ inherited from 
the steady flow solution. It appeared that the evolution did not produce any noticeable changes; 
this confirms the validity of the assumption of stationarity. Next, we introduce an additional perturbation. 
In our case it was applied to the velocity $v^R$ (the other three initial data $m(R)$, $n(R)$, and $\epsilon(R)$ 
are the same as in the steady flow); $v^R$ was perturbed by a bell-shape profile of a sine wave restricted 
to one half of its period. Such perturbation, initially located outside the sonic radius, produces two signals: 
one traveling outwards and one towards the center. If the initial amplitude of the perturbation is very small 
and its support is relatively narrow, the signal propagating inwards passes through the sonic point 
and eventually disappears through the inner boundary (it falls onto the central body).

\begin{figure}
\begin{center}
\includegraphics[width=80mm]{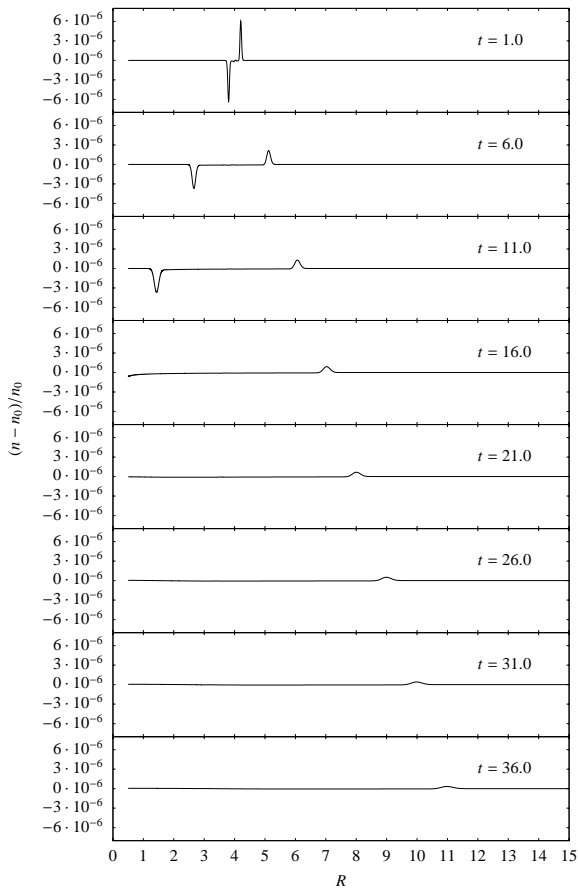}
\end{center}
\caption{Evolution of small perturbations. Here $n_0$ denotes the density in the unperturbed flow. 
The snapshots show the density contrast $(n - n_0)/n_0$ in the chronological order.}
\label{small_perturbations}
\end{figure}

Fig.\,\ref{small_perturbations} shows the evolution of tiny perturbations of compact support. Here, 
in order to visualize any changes, we were forced to plot the contrast of density, i.e.,  $(n - n_0)/n_0$, 
where $n_0$ refers to the unperturbed flow. In our case this value is of order $10^{-6}$, 
demonstrating the quality of the method by being able to handle solutions up to such precision. 
This, however, requires some fine tuning in the post-processing of the data. It is, for instance, 
well known that the very process of interpolating the initial solution onto a new grid introduces 
small numerical errors, which can easily be observed at high precision. Thus, in order to eliminate 
such effects, we have evolved the unperturbed flow to get the values of $n_0$ used to calculate 
the density contrast at different times.

The background solution used in this example corresponded to the following parameters: 
The exponent in the polytropic equation of state was set to $\Gamma = 1.4$, the asymptotic mass 
of the whole configuration has been normalized to unity, the outer boundary of the cloud was
 placed at $R_\infty = 10^6$, the asymptotic baryonic density was equal $n_\infty = 1.6 \cdot 10^{-19}$, 
and the asymptotic sound velocity was set to $a_\infty^2 = 0.1$ (here and in all other results 
we have adopted the gravitational units with $c = G = 1$). For such a solution the sonic point 
and the apparent horizon were located at $R_\ast = 0.82$ and $R_\mathrm{BH} = 0.34$ respectively, 
and the mass contained in the fluid constituted the bulk of the entire mass, namely $m_\mathrm{fluid} = 0.83$. 
The accretion rate for this solution reaches a very small value of $\dot m = 2.6 \cdot 10^{-18}$, 
thus the growth of the central object can be neglected during entire simulation (this fact has been 
confirmed independently by allowing the central mass to grow according to the actual value of $\dot m$).

\section{Sonic horizons versus apparent horizons}

For large initial perturbations the situation can be more complex, as will be discussed below. 
A discontinuous solution with shocks can develop, and we can observe some reflection of 
the signal that was initially propagating inwards. 

\begin{figure}
\begin{center}
\includegraphics[width=80mm]{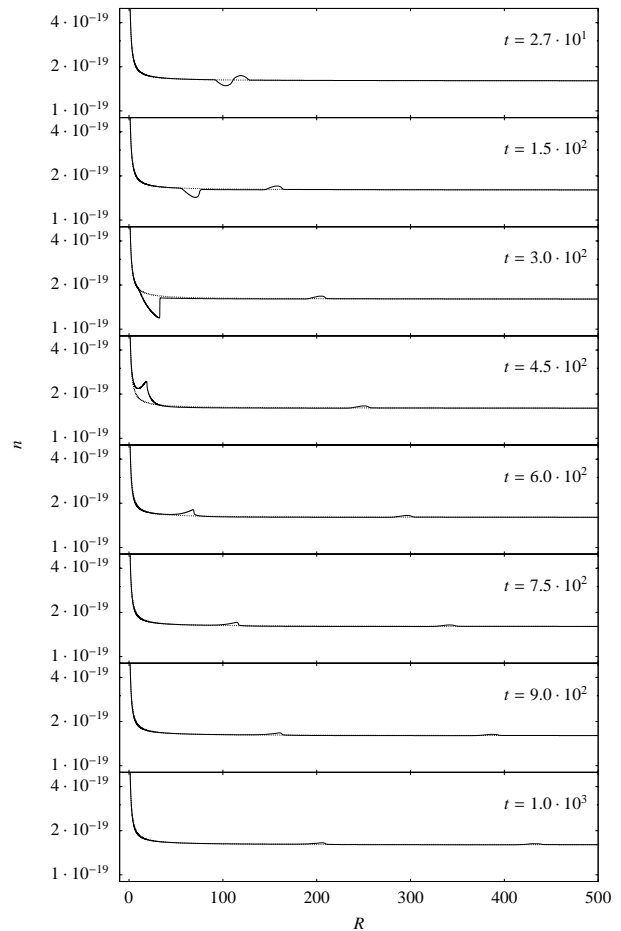}
\end{center}
\caption{Evolution of the perturbed density. The snapshots are placed in the chronological order. 
The profile corresponding to the unperturbed flow is depicted with a dotted line.}
\label{density_evolution}
\end{figure}

Fig.\,\ref{density_evolution} shows snapshots from the evolution of the perturbations applied to the 
same background solution as before. In each of the graphs on Fig.\,\ref{density_evolution} the perturbed 
density profile has been plotted over the profile corresponding to the steady solution. One can clearly 
see the stable behavior on these plots even though they are limited to the small radius range of $R < 0.5 
\cdot 10^3$ and relatively short evolution times. The original simulations have been performed during much longer 
times confirming the stability of the flow. At late stages of the evolution the initial perturbation 
was indistinguishable from the numerical noise.

All these numerical results suggest the stability of the accretion flow also in the regime where the
 mass of the fluid is large. Naturally, numerical simulations of this kind can never replace a strict
 mathematical proof, because we cannot investigate the whole family of possible initial data (meaning 
both steady solutions and perturbation profiles).
 
\begin{figure}
\begin{center}
\includegraphics[width=80mm]{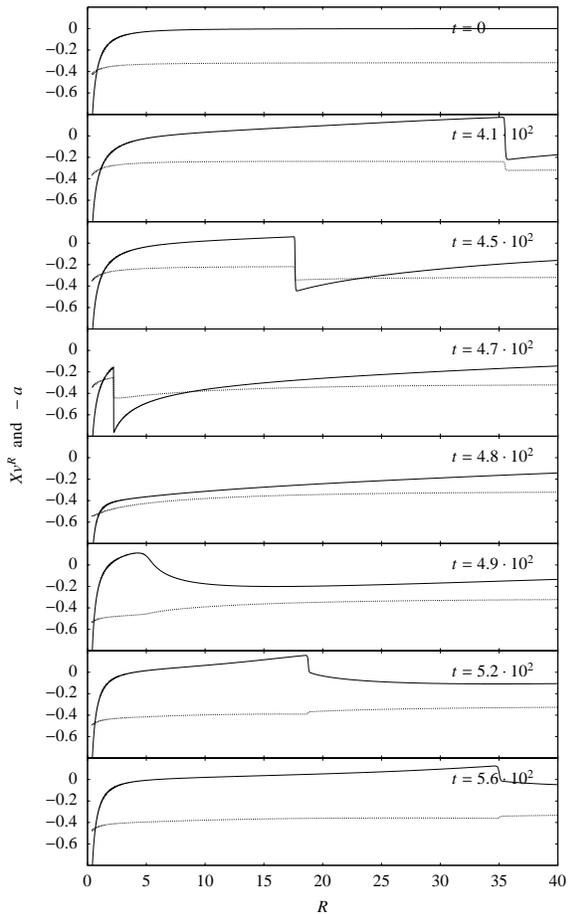}
\end{center}
\caption{Reflection of the incoming signal from the inner parts of the accretion cloud. The velocity $Xv^R$ 
is drawn with the solid line, while the dotted line depicts values of $-a$. Intersections of the two graphs 
correspond to the sonic points (see the discussion in text).}
\label{reflection}
\end{figure}
 
The reflection of the strong incoming signal from the inner parts of the accreting cloud is somewhat surprising. 
For a small and compact perturbation entering the supersonic region in the center of the cloud it should not 
be possible to be reflected and reach the subsonic region again---this would require the perturbation to 
travel with a velocity which, relative to the unperturbed flow, should be greater than the speed of sound. 
Due to that fact, in analogy with the black hole horizon, the term ``sonic horizon'' has been coined to name 
the surface bounding the supersonic region.

A careful inspection reveals that, contrary to the event horizons surrounding black holes, this notion can only 
be approximate. For strong nonlinear perturbations the concept of an unperturbed background solution loses meaning.
 The values of the local sound speed $a$ change due to the perturbation and new sonic points can appear. 
In addition, the speed of a strong shock wave is no longer limited to the local speed of sound.

Such behavior has been illustrated on Fig.\,\ref{reflection}. Here we are still dealing with the same 
initial solution,  perturbed slightly more to show the whole phenomenon in a clearer way. 
The quantity $Xv^R$ is plotted with the solid line, while for the graph of $-a$ a dotted line has been used. 
A part of the initial perturbation develops into the shock propagating inwards which, after some time of 
evolution, creates additional sonic points (i.e., intersections of graphs of $Xv^R$ and $-a$) lying outside 
the original ``sonic horizon.'' When the perturbation reaches the inner parts of the cloud it even destroys 
the ``original'' sonic point, which is being replaced by the newly created one. The reflected signal really 
becomes visible in the broad structure outside the new sonic point, or to say this differently, 
the outermost sonic point moves toward the center releasing the outgoing perturbation, which 
then propagates freely outwards.

The behavior of small  perturbations of compact support confirms the standard interpretation 
of sonic horizons as analogs of event horizons in the linear regime. However, results concerning 
strong perturbations suggest that one should be careful in  formulating the analogy (quite common 
in the so-called analogous gravity models) between the sonic horizon and the event horizon \cite{das, dasgupta}.
 
In summary, our numerical results reported in the last two sections suggest that in both cases---of small 
and large perturbations---steady accretion is stable. The amplitudes of the perturbations decrease, and 
after sufficient time we are left with the background solution. This can be demonstrated for both 
accretion regimes: solutions with a large mass in the center and having little fluid and systems 
with large amount of gas but possessing light compact cores.


\section{Stability of the Newtonian accretion}

\begin{figure}[t!]
\begin{center}
\vspace{9cm}
\vspace{-1.2 cm}
{\includegraphics[width=80mm]{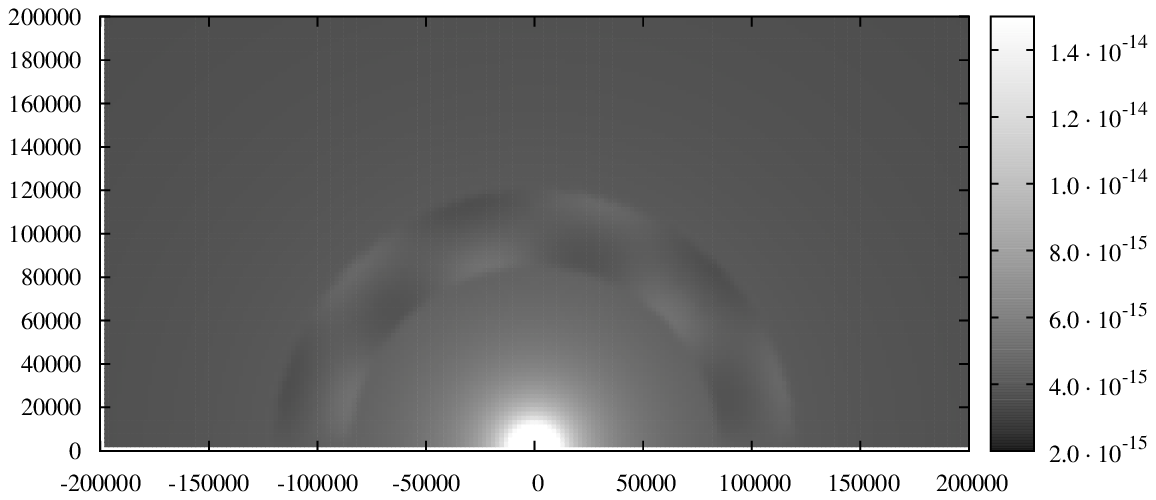}\par}
\vspace{-2.4 cm}
{\includegraphics[width=80mm]{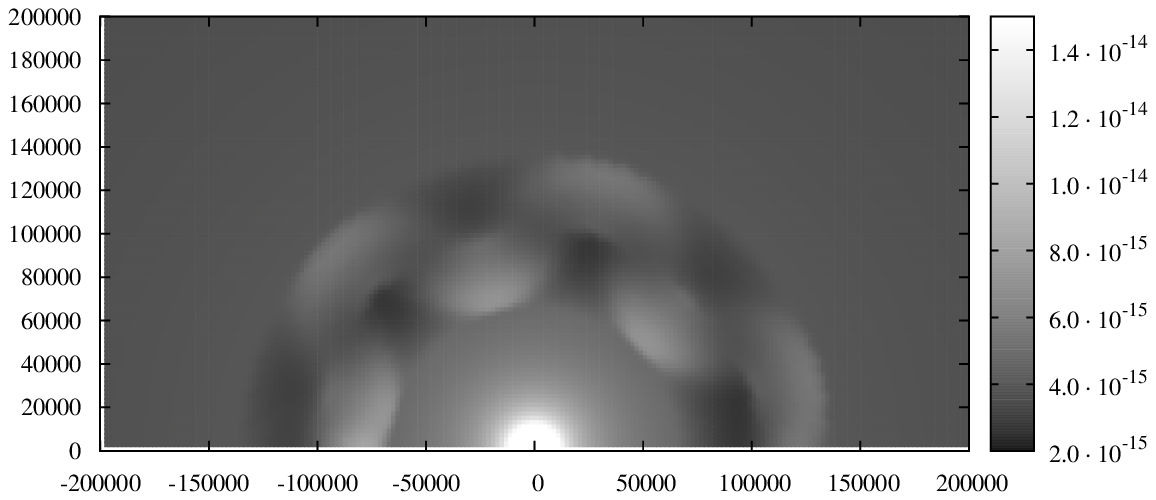}\par}
\vspace{-2.4 cm}
{\includegraphics[width=80mm]{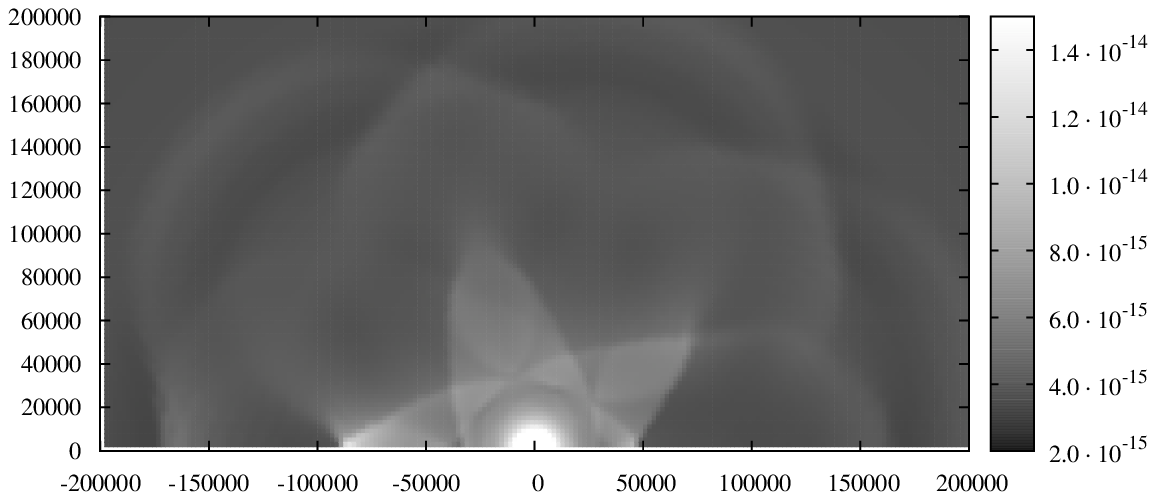}}
\end{center}
\vspace{-1.2 cm}
\caption{Evolution of axially symmetric density perturbations of the steady accretion cloud. The plots 
show the spatial distribution of the density at times $5 \cdot 10^3$, $5 \cdot 10^4$, and $2.5 \cdot 10^5$ respectively.}
\label{axi_1}
\end{figure}

In this Section we shall report results obtained with the use of a version of the \textsc{Prometheus} 
code by Fryxell, M\"{u}ller, and Arnett \cite{fryxell_mueller}, adapted to our purposes. \textsc{Prometheus} 
is a Newtonian HRSC hydrodynamical code implementing the original PPM reconstruction scheme developed by 
Colella and Woodward \cite{colella_woodward}. It has been extensively used for simulating such astrophysical
 phenomena as supernova explosions and it is capable of simulating self-gravitating flows both in spherically 
and axially symmetric cases.
As the initial data we have used solutions to the Newtonian equations for the steady flow that include 
self-gravitation of the accreting fluid. An equation of state was polytropic, $p = K \rho^\Gamma$. 
A point mass $m_\mathrm{P}$ was assumed to exist at $R = 0$.
 
Results concerning spherically symmetric perturbations have been already presented in \cite{zakopane}. 
They are qualitatively similar to the relativistic results reported in the preceding sections. 
The accreting flow was stable both in the test and fluid-rich regimes. Large incoming perturbations 
were being reflected from the inner parts of the accretion cloud. We have also observed effects 
analogous to those described in the previous chapter concerning the creation and destruction of the sonic 
horizons. Moreover, for sufficiently small perturbations of compact support no reflection has been noticed.

\begin{figure}[t!]
\begin{center}
\vspace{-1.2 cm}
{\includegraphics[width=80mm]{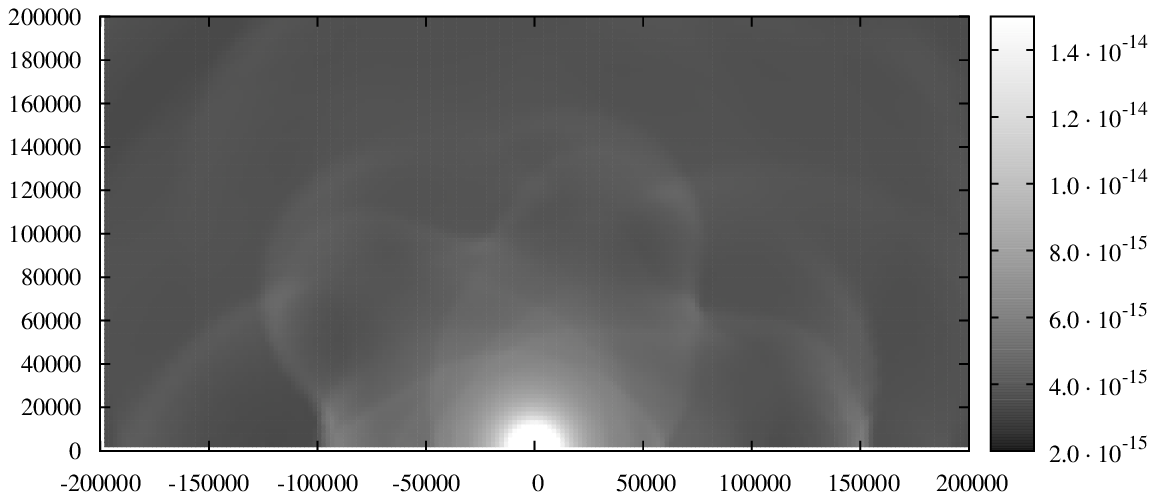}\par}
\vspace{-2.4 cm}
{\includegraphics[width=80mm]{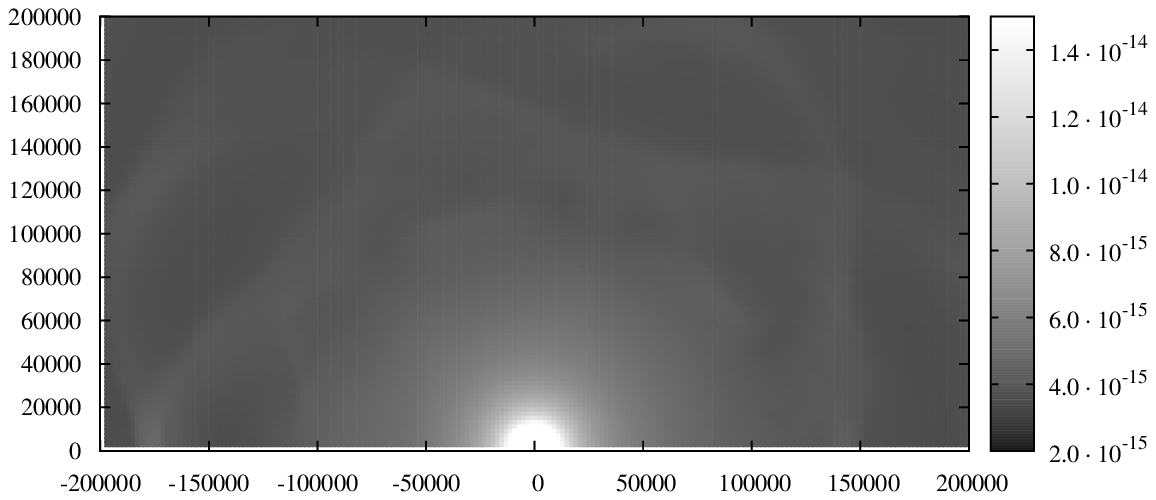}\par}
\vspace{-2.4 cm}
{\includegraphics[width=80mm]{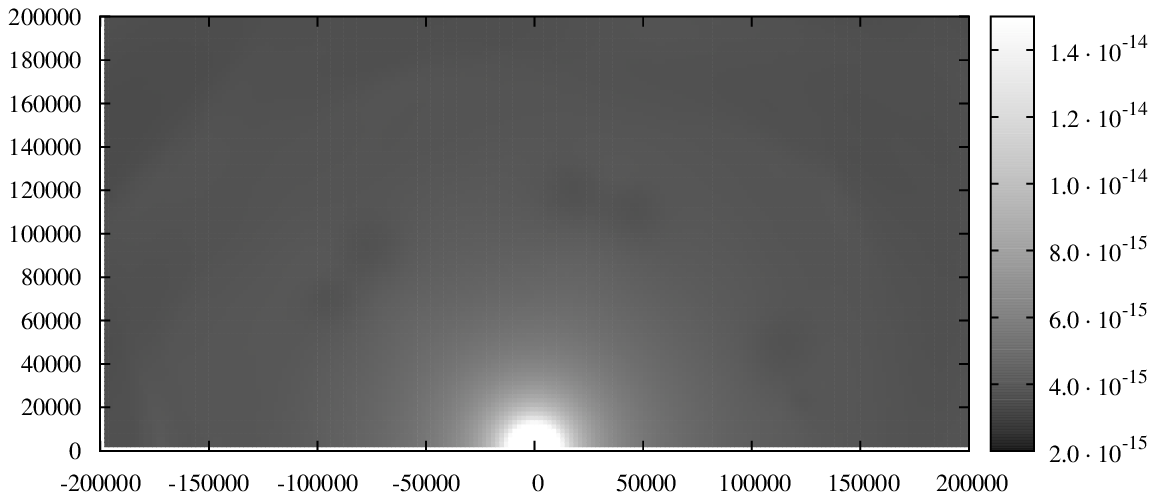}}
\end{center}
\vspace{-1.2 cm}
\caption{Continuation of the previous figure. The subsequent snapshots correspond to evolution times 
of $4.5 \cdot 10^5$, $6.5 \cdot 10^5$, and $8.5 \cdot 10^5$ respectively.}
\label{axi_2}
\end{figure}

Below we shall report studies of axially symmetric perturbations. In \textsc{Prometheus} 
simulations of two or three dimensional flows are 
being performed using the so-called dimensional splitting. The local Riemann problems appearing in 
the Godunov-type method are being solved separately in each of the dimensions, but, in order to 
preserve consistency of the method, every time-step the transversal components of velocity are 
also being evolved, using the effective advection equations. The subtle point about this 
procedure is that the order of the subsequent one dimensional sweeps is important to provide 
the desired accuracy of the whole scheme. This issue has been discussed in detail by Strang i
n \cite{strang}.

The gravitational potential has to be found at each time-step  by solving the gravitational 
Poisson equation. The \textsc{Prometheus} code implements a method, based on the expansion 
of the gravitational potential in terms of the spherical harmonics, that has been developed 
by M\"{u}ller and Steinmetz \cite{mueller_steinmetz}.

In our case the Euler equations of hydrodynamics and the Poisson equation for the gravitational 
field were solved on a spherical grid consisting of 600 zones in the radial and 180 zones 
in the angular direction.

The results presented on Fig.\,\ref{axi_1} have been obtained for the following parameters of the flow. 
The outer boundary of the cloud was assumed to be located at $R_\infty = 2 \cdot 10^6$, 
the central mass was set to $m_\mathrm{P} = 3 \cdot 10^3$ (in the Newtonian limit gravitational 
units that we use here correspond to setting $G = 1$). As before, we have chosen a solution 
where most of the mass is contained in the fluid, i.e., $m_\mathrm{P}/m_\mathrm{tot} = 3\%$. 
The asymptotic parameters of the unperturbed flow were as follows: $\rho_\infty = 3 \cdot 10^{-15}$, 
$U_\infty = -4 \cdot 10^{-5}$, and for the parameters of the polytropic equation of state we have 
taken $\Gamma = 1.4$, $K = 5 \cdot 10^4$.

Initial data coincide with the relevant characteristics of the stationary flow, with the exception of the velocity. 
The velocity perturbations are of the form
\[ \delta U (R, \theta) = \left\{ \begin{array}{ccl} A \sin \left( \frac{R - R_a}{R_b - R_a} \pi \right) \sin (6 \theta) 
& 
\mathrm{for} & R_a < R < R_b, \\ 0 & \mathrm{for} & R < R_a \;\; \mathrm{or} \;\;  R_b < R, \end{array} \right. \]
where $A$  is the initial amplitude of the perturbations while $R_a$ and $R_b$ describe their support. 
This is obviously a completely arbitrary choice. The only important feature of these perturbations 
is their nonsphericality.
 
The boundary conditions have been chosen in the same way as in the relativistic case. 
At the inner boundary the outflow conditions were assumed, while at the outer boundary 
we have implemented ``external'' inflow conditions based on the background steady flow. 
Fig.\,\ref{axi_1} and \ref{axi_2} show snapshots from the evolution of the density, 
which has been color-coded using a logarithmic scale.

As expected, after a sufficient time all perturbations either disappear through 
the inner boundary or disperse outwards. This again suggests that the accretion flow 
is stable, even in the regime where the mass of the fluid dominates over that of the 
central object. In particular, we do not observe any fragmentation of the cloud, which  
might be expected  in the case of non-spherical, self-gravitating accretion flows. 
A simple theoretical argument, basing on the unproven \cite{bonnor}  but unreasonably 
effective Jeans criterion, goes as follows. The Jeans length can be estimated as 
$R_\mathrm{J} = a_\infty/\sqrt{\varrho_\infty} \approx 6 \cdot 10^6$. This gives 
a value three times larger than the size of the entire accreting cloud, i.e., 
$R_\infty = 2 \cdot 10^6$. That hints at the absence 
of a fragmentation due to self-gravity, which agrees with our numerical results. On the
other hand it is not clear whether the Jeans
criterion can be applied to general    nonspherical perturbations and the instability in the
general case cannot be excluded.
 
\section{Conclusions}

This paper is dedicated to the discussion of the stability of steady accretion of perfect fluids onto compact objects,
 with an emphasis on the self-gravitation of the accreting gas. The recently discovered steady accretion flows are rich
 in the fluid and the backreaction effects are important.  Known analytic results do not apply to such systems. 
The stability investigation requires a proper handling of a set of nonlinear partial differential equations and 
it is accessible only by means of numerical computations. Such a numerical analysis has been performed in preceding 
Sections using modern, high resolution and shock-capturing schemes, both in the general-relativistic case as well 
as in Newtonian hydrodynamics. In all examined cases the flows have been stable, even for large nonlinear perturbations.
Simulations of the evolution of large, sometimes even discontinuous, perturbations have led to a remarkable observation 
on the so-called ``sonic horizons.'' In the linear regime, with very small perturbations, the sonic horizon can be 
viewed as a surface bounding a region from which no perturbation can escape; it is impenetrable from inside, 
analogous to the event horizons in general relativity. We show that for strong perturbations this 
is no longer true. Perturbations  can  change positions of the sonic points and easily escape from a region 
initially bounded by the sonic horizon. Thus, the analogy between the ``sonic horizon'' and the event horizon 
of a black hole is rather limited.
 
Our analysis of the stability is restricted to the accretion cloud only. In all numerical simulations we had to 
ensure that the gas is being delivered to the system with a small, constant rate. To relax this assumption one 
would have to take into account a physical process that could be responsible for the feeding of the accretion cloud. 
One of the possibilities is to consider the so-called ``quasistars'' \cite{begelman_et_al}. They consist 
of the spherical accretion cloud surrounded by a large, highly massive stellar envelope, being a reservoir of 
gas necessary to support the accretion. The other class of objects might constitute Thorne--\.Zytkow stars 
\cite{thorne,thorne2}.

In section IV we described a number of classical tests for numerical codes. Stationary accretion flows posess 
all essential elements---velocity field, selfgravitation and pressure. These solutions are stable. We think 
that they should be included into the standard test suite for general-relativistic hydrodynamical codes.

Spherically symmetric models of accretion must be understood as a highly idealized version of physical 
reality. In most astrophysical scenarios the inflowing gas is observed in the form of accretion discs deviating
 strongly from spherical symmetry. Nevertheless, models of spherical accretion occupy an important place 
in the theoretical astrophysics as elements of a more complex description of astrophysical phenomena 
(see, e.g., the aforementioned works on ``quasistars'' by Begelman, Rossi, and Armitage \cite{begelman_et_al}). 
They allow for the inspection of the effects caused by self-gravity in general-relativistic hydrodynamics 
in a simple, but nontrivial, case.

More realistic models should take into account the radiation originated and transported through the accretion 
cloud. Recently a Newtonian analysis of such processes has been published by Karkowski, 
Malec, and Roszkowski \cite{karkowski_malec_roszkowski}. These works are being continued in 
the general-relativistic context, the results revealing the importance of self-gravity and 
its connection to the observational characteristics of the model such as the total luminosity 
of the accretion cloud or the redshift of the emitted radiation.

Acknowledgments. This paper has been partially supported by the MNII grant 1PO3B 01229. Numerical 
computations have been made at the Academic Computer Center Cyfronet, grant MNiSW/SGI3700/UJ/116/2007. 
PM thanks Ewald M\"{u}ller
for kind agreement to use the Prometheus code and for the possibility to visit the Institute of 
Astrophysics at Garching. Authors thank Niall \'O Murchadha for careful reading and useful comments.


\begin{thebibliography}{aa}
\bibitem{bondi_hoyle} H. Bondi, F. Hoyle, \textit{Mon. Not. R. Astron. Soc.} \textbf{104}, 273 (1944).
\bibitem{hoyle_lyttleton} F. Hoyle, R.A. Lyttleton, \textit{Proc. Cam. Phil. Soc.} \textbf{35}, 405 (1939).
\bibitem{lyttleton_hoyle} R.A. Lyttleton, F. Hoyle, \textit{The Observatory}, \textbf{63}, 39 (1940).
\bibitem{bondi} H. Bondi, \textit{Mon. Not. R. Astron. Soc.} \textbf{112}, 195 (1952).
\bibitem{michel}F.C. Michel, \textit{Astrophys. Space Sci.} \textbf{15}, 153 (1972).
\bibitem{malec} E. Malec, \textit{Phys. Rev.} \textbf{D60}, 104043 (1999).
\bibitem{kkmms} J. Karkowski, B. Kinasiewicz, P. Mach, E. Malec, Z. \'{S}wierczy\'{n}ski, \textit{Phys. Rev.} 
\textbf{D73}, 021503(R) (2006).
\bibitem{thorne} K.S. Thorne, A.N. \.{Z}ytkow, \textit{Astrophys. J.} \textbf{199}, L19 (1975).
\bibitem{thorne2} K.S. Thorne, A.N. \.{Z}ytkow, \textit{Astrophys. J.} \textbf{212}, 832 (1977).
\bibitem{begelman_et_al} M.C. Begelman, E.M. Rossi, Ph.J. Armitage, \textit{Mon. Not. R. Astron. Soc.} in press.
\bibitem{papadopoulos_font} P. Papadopoulos, J. Font, \textit{Phys. Rev.} \textbf{D61}, 024015 (1999).
\bibitem{Zampieri} L. Zampieri, J.C. Miller and R. Turolla, \textit{Mon. Not. R. Astron. Soc.} \textbf{281}, 1183 (1996).
\bibitem{Nobili} L. Nobili, R. Turolla, L. Zampieri \textit{Astrophys. J.} \textbf{383}, 250 (1991).
\bibitem{moncrief} V. Moncrief, \textit{Astrophys. J.} \textbf{235}, 1038 (1980).
\bibitem{ladek} B. Kinasiewicz, P. Mach, E. Malec, \textit{International Journal of Geometric Methods in Modern Physics}
 \textbf{4}, 197 (2007).
\bibitem{romero_1996} J.V. Romero, J.$\mathrm{M^{\underline{a}}}$ Ib\'{a}\~{n}ez, J.$\mathrm{M^{\underline{a}}}$ 
Mart\'{\i}, J. Miralles, \textit{Astrophys. J.} \textbf{462}, 839 (1996).
\bibitem{banyuls_et_al} F. Banyuls, J.A. Font, J.$\mathrm{M^{\underline{a}}}$ Ib\'{a}\~{n}ez,
 J.$\mathrm{M^{\underline{a}}}$ Mart\'{\i}, J.A. Miralles, \textit{Astrophys. J.} \textbf{476}, 221 (1997).
\bibitem{van_leer} B. Van Leer, \textit{J. Comput. Phys.} \textbf{32}, 101 (1979).
\bibitem{donat_maraquina} R. Donat, A. Maraquina, \textit{J. Comput. Phys.} \textbf{125}, 42 (1996).
\bibitem{genesis} M.A. Aloy, J.$\mathrm{M^{\underline{a}}}$ Ib\'{a}\~{n}ez, J.$\mathrm{M^{\underline{a}}}$ Mart\'{\i}, 
E. M\"{u}ller, \textit{Astrophys. J. Suppl.} \textbf{122}, 151 (1999). 
\bibitem{gourgoulhon} E. Gourgoulhon, \textit{Astron. Astrophys.} \textbf{252}, 651 (1991).
\bibitem{blandford} R.D. Blandford, C.F. McKee, \textit{Physics of Fluids} \textbf{19}, 1130 (1976).
\bibitem{tooper} R.F. Tooper, \textit{Astrophys. J.} \textbf{142}, 1541 (1965).
\bibitem{hairer} E. Hairer, S.P. Norsett, G. Wanner, \textit{Solving ordinary differential equations I.
 Nonstiff problems}, Springer Series in Computational Mathematics, Springer--Verlag (1993).
\bibitem{brown_et_al} P.N. Brown, A.C. Hindmarsh, L.R. Petzold, \textit{SIAM J. Sci. Comput.} \textbf{15}, 1467 (1994).
\bibitem{li_petzold} S. Li, L.R. Petzold, \textit{Design of new DASPK for sensivity analysis,} 
Technical Raport, Department of Computer Science, Universit of California Santa Barbara (1999).
\bibitem{karkowski_malec_roszkowski} J. Karkowski, E. Malec, K. Roszkowski, \textit{Astron. Astrophys.}
 \textbf{479}, No. 1, 167(2008)
\bibitem{das} T.K. Das, \textit{Class. Quant. Grav.} \textbf{22}, 2971 (2005).
\bibitem{dasgupta} S. Dasgupta, N. Bili\'{c}, T.K. Das, \textit{Gen. Rel. Grav.} \textbf{37}, 1877 (2005).
\bibitem{fryxell_mueller} B.A Fryxell, E. M\"{u}ller, W.D. Arnett, Max-Planck-Institut f\"{u}r Astrophysik,
 Garching, Preprint 449 (1989).
\bibitem{colella_woodward} Ph. Colella, P.R. Woodward, \textit{J. Comput. Phys.} \textbf{54}, 174 (1984).
\bibitem{zakopane} P. Mach, \textit{Acta Phys. Pol.} \textbf{B38}, 3935 (2007).
\bibitem{strang} G. Strang, \textit{SIAM J. Numer. Anal.} \textbf{5}, 506 (1968).
\bibitem{mueller_steinmetz} E. M\"{u}ller, M. Steinmetz, \textit{Comput. Phys. Commun.} \textbf{89}, 45 (1995).
\bibitem{bonnor} W.B. Bonnor, \textit{Mon. Not. R. Astron. Soc.} 117, 104(1957).
 \end{thebibliography}
\end{document}